# Dual origin of $E=mc^2$

Jean-Paul Auffray

*07700 Saint-Martin d'Ardèche, France*
(e-mail: jpauffray@yahoo.fr)

**Abstract**. The mass-energy relation $E=mc^2$ has a dual origin, one grounded in the postulate of the existence of an aether made of "ultramondane particles" moving in space at the speed of light, $c$; the other, a consequence, first deduced by Henri Poincaré, of John Poynting's celebrated electromagnetic Theorem.
  **Keywords:** Mass-energy, Special relativity, Poynting, Poincaré, Einstein.

## Introduction

In a *Brief communication* recently published in *Nature* [1], Simon Rainville *et al.* present experimental evidence which indicates that the mass-energy relation $E=mc^2$ "holds to a level of at least 0.00004%", a most remarkable result. The communication opening sentence reflects generally accepted views concerning the origin of the famous equation. The authors assert: "One of the most striking predictions of Einstein's special theory of relativity is also perhaps the best known formula in all of science $E=mc^2$."

We wish to show that this statement is incorrect on several counts. This is of importance in view of the second proposition presented in the above-cited communication: "Il this equation were found to be even slightly incorrect, the impact would be enormous – given the degree in which special relativity is woven into the theoretical fabric of modern physics and into everyday applications such as global positioning systems."

## 1. Origins of $E=mc^2$

Newton's 1704 Query: "Are not gross Bodies and Light convertible into one another, and may not Bodies receive much of their Activity from the Particles of Light which enter their Composition?"[2] is generally regarded as the first modern suggestion of an inherent relationship between mass and "Activity" (Energy). Newton did not quantify his proposal, however. $E=mc^2$ was subsequently established as such as the results of two drastically different lines of thought.

The first line originated in Le Sage's 1784 theory of gravitation grounded in the postulate of the existence of an aether made of "ultramondane particles" moving in space at the speed of light, $c$. This led English philosopher and theologian Samuel Tolver Preston to express the idea in 1875 that aether is "rarefied mass" and mass "concentrated aether" [3], a view which received a profound echo in Cambridge, with sir Joseph John Thomson and the members of the Society for Psychical Research, and in France, in the mind of College de France professor Paul Langevin, who held that matter is "liquefied aether".

According to Leibniz's 1686 fundamental concept of an inherent *vis motrix* equal to $mv^2$ for a body of mass $m$ moving at the velocity $v$, Samuel Preston found, for $v=c$, an energy equal to $mc^2$. This line of thought was revived in 1903 when, encouraged by his friend, the famed astronomer Giovanni Schiaparelli, Italian railroad engineer Olinto de Pretto published in Venice the extensive memoir [4] that modern tenants of this line of thoughts regard as the first legitimate formulation of $E=mc^2$.

The second line of thought seeking to establish the mass-energy relation has its roots in the electromagnetic theory. If $E$ designates a given amount of electromagnetic or *radiation* energy, then $E=mc^2$ associates the mass $m=E/c^2$ to this energy. Indeed, this is how, using arguments not based on the aether postulate, $E=mc^2$ was first established. As pointed out by Albert Einstein in 1906 [5], although several experts in the electromagnetic theory flirted with the discovery prior to 1905 [6] the equation was first formerly extracted from the electromagnetic theory through a reasoning constructed by French mathematician Henri Poincaré in 1900.

Poincaré's publication precedes the invention of special relativity by five years and it is therefore incorrect to view $E=mc^2$, at least when written as $m=E/c^2$, as a "prediction" of special relativity. We briefly review how Poincaré established $m=E/c^2$ in 1900 [7].

## 2. Henri Poincaré's 1900 discovery of $E=mc^2$

After James Maxwell's untimely death en 1879 at the age of forty-eight, his former student John Poynting, then physics professor at Mason Science College, future Birmingham University, constructed the telling representation of the electromagnetic theory which became known as the "Poynting's Theorem". One of the first on the continent to understand the importance of Poynting's work was Poincaré. Using Poynting's Theorem, he obtained $m=E/c^2$ quite simply. Briefly summarized and expressed in modern notation, his calculation runs as follows:

If $c$ designates the speed of light in vacuum and $E$ the electromagnetic energy, then the radiation energy flux is $S=Ec$. Poincaré now introduced a strong innovation: he assumed that the emitted radiation carries a momentum, $p$. By Poynting's Theorem, this momentum must be equal to $S/c^2$. By Descartes's formula, one must also have $p=mv$. Setting $v=c$, since the emitted radiation travels at the speed of light, and combining $p=mc=S/c^2$ with $S=Ec$, Poincaré readily obtained a result he expressed in these terms: "We can regard electromagnetic energy as a fictitious fluid of density $K_0 J$ which moves in space in accordance with Poynting's laws."

To fully appreciate the meaning of this statement, let us translate it in modern terms.

In Section §627 of his *Treatise on Electricity & Magnetism*, Maxwell assigns the symbol $K$ to the quantity Michael Faraday had called Specific Inductive Capacity or Dielectric Constant before him. Maxwell noted that $K$ has the dimensions of the inverse of a velocity squared. If $c$ designates the speed of light and if $K_0$ is the value of the Dielectric Constant in vacuum, then $K_0=1/c^2$. In his 1900 memoir, Poincaré assigns to the electromagnetic energy the symbol $J$ (rather than $E$, as we do today). Since, in this context, "density" means "mass" and $K_0J=E/c^2$ as indicated above, Poincaré's 1900 statement translates into $m=E/c^2$.

In a lecture he delivered at the Sorbonne in 1899, Poincaré expressed his discovery in a way which illustrates the fact that his primary concern at this stage of his research was rooted in the principle of momentum conservation: "If any apparatus whatever, after having produced electromagnetic energy, sends it by radiation in a certain direction, wrote Poincaré, the apparatus must recoil like a cannon which has launched a projectile."

He gave a numerical illustration of the phenomenon: "If the apparatus has a mass of one kilogram, and if it has sent in one direction, with the velocity of light, three millions of joules, the velocity due to the recoil is 1 cm per second." [8].

It remained to show that $m=E/c^2$ also applies the other way around, i.e. when $m$ designates an *inertial* mass associated with matter. In his 1900 paper, Poincaré explicitly states that the total mass of the apparatus is equal to the material mass plus the mass carried off as $E/c^2$ by the electromagnetic "bullet" the apparatus has emitted. Recalling that in 1900 physicists universally held true Lavoisier's principle of mass conservation, Poincaré's statement may be construed to mean that, after emission, the electromagnetic pulse having carried off a mass equal to $E/c^2$, the apparatus material mass must have decreased concurrently by an equivalent amount – a conclusion that Poincaré did not state explicitly in these terms in 1900, however; but he did not stop there. Five years later, he published in the *Rendiconti del Circolo Matematico di Palermo* his momentous memoir *Sur la dynamique de l'électron* [9], a much extended version of a short Note (five pages) he had published the previous month in *Comptes rendus de l'Académie des sciences de Paris* [10]. He formally presents in this memoir a new (electro)dynamics compatible with the "Principe de relativité" he had formulated earlier. He named the new scheme *Mécanique nouvelle*. We call it today Relativistic Mechanics. In Section §8 of the memoir, Poincaré derives an expression for the velocity dependence of the electron mass. When expressed in modern notation, his result states that if the electron velocity is $v$ and its "rest mass" is $m_0$ (the mass when $v=0$), then, for velocities much less than the velocity of light, the electron total energy (potential plus kinetic) is $m_0c^2+(1/2)m_0v^2$. This telling result means that the electron rest mass corresponds to an inherent reserve of energy equal to $m_0c^2$. This formally establishes $E=mc^2$ on the basis of relativistic considerations.

In the meantime, Fritz Hasenöhrl had published his important thermodynamically inspired paper [11] in which he shows that to the "mechanical mass" of a hollow enclosure filled with radiation must be added an "apparent mass" equal to $8E/3c^2$, a result he later [12] corrected to $4E/3c^2$.

## 3. In 1905, Albert Einstein did not properly derive $E=mc^2$

Einstein sent his first paper concerning $E=mc^2$ to *Annalen der Physik* in September 1905 [13], several weeks after Poincaré's memoir had been received in Palerme. In it, he attempted, but failed (see below), to prove the validity of $E=mc^2$ using a relativistic line of reasoning. In this paper, Einstein uses Poincaré's model of an apparatus which emits electromagnetic radiation; but while Poincarè's emitter operates in a single direction, Einstein's apparatus emits equal amounts of energy in two opposite directions, thus avoiding the Poincaré recoil.

The body which emits energy is first observed from a platform in which it is at rest. Its energy before emission in this case is E. It is also observed from a platform which moves uniformly with respect to the body. Observed from this platform, the body energy (before emission) is H. At a crucial point in his reasoning, Einstein skips a line, claiming: "It is clear that the difference $H – E$ can differ from the kinetic energy $K$ of the body with respect to the other system only by an additional constant $C$." Consequently he wrote $H – E = K + C$.

It has been shown since that this claim presupposes the very relation $E=mc^2$ it seeks to establish. In brief, Einstein's 1905 reasoning seeking to establish $E=mc^2$ on the basis of the model he proposed is circular [14]. In all events, his conclusion: "If the theory agrees with the facts, then radiation transmits inertia between emitting and absorbing bodies" is a straightforward restatement of Poincaré's 1900 pre-relativistic fundamental result.

## Conclusion

The discovery of the mass-energy relation $E=mc^2$ cannot properly be attributed to Albert Einstein [15]. Einstein's world-wide notoriety has precluded general recognition and acceptance of the dual origin and true nature of $E=mc^2$.

## References

1. Rainville, S. *et al. Nature* **438**, 1096-1097 (2005).
2. Newton, I. *Opticks*, Query 30 (1704).
3. Preston, S. T. *The Physics of Aether* (1875).
4. De Pretto, O. *Reale Istituto Veneto di Scienze, Lettere ed Arti,* **LXIII**, II, 439-500 (1904).

5. Einstein, A. *Ann. der Phys.* **20**, 627-633 (1906).
6. Notably J. J. Thomson (1881) and Oliver Heaviside (1889).
7. Poincaré, H. *Arch. néerland. sci.* **2**, 5, 252-278 (1900).
8. Poincaré, H. *Electricité et Optique*, §351 (1901).
9. Poincaré, H. *Rend. Circ. mat. Palermo* **21**, 129-176 (1906).
10. Poincaré, H. *Compt. Rend.* **140**, 1504-1508 (1905).
11. Hasenörhl, F. *Wien Sitz.* **IIa**, 113, 1039 (1904); *Ann. der Phys.* **15**, 344 (1904).
12. Hasenörhl, F. *Ann. der Phys.* **16**, 589 (1905). Later, Max Planck (1907), Paul Langevin (1913) and others provided conditions for possible energy release in the destruction of mass, as in fission or fusion of atomic nuclei.
13. Einstein, *A. Ann. der Phys.* **18**, 639-641 (1905).
14. Ives, H. E. *J. Opt. Soc. Amer.* **42**, 8, 540-542 (1952).
15. J. Stachel and R. Torreti have attempted to rehabilitate Einstein's 1905 and 1906 lines of reasoning. Cf. Stachel, J. and Torreti, R., *Am. J. Phys.* **50**, 8, 160-761 (1982). But Ives's conclusions have been confirmed by several authors since. See for instance A. A. Logunov, *Henri Poincaré and Relativity Theory,* Nauka, (2005), p. 124.